\documentstyle[amssymb,preprint,aps]{revtex}
\begin{document}
\draft
\title{Vortex-antivortex configurations and their stability in mesoscopic
superconducting square.}
\author{$^{a,b}$T. Mertelj and $^a$V.V. Kabanov}
\address{$^a$J. Stefan Institute 1001, Ljubljana, and\\
$^{b}$Faculty of Mathematics and Physics, University of Ljubljana,\\
Ljubljana, Slovenia}
\date{\today}
\maketitle

\begin{abstract}
\widetext
We solve the Ginzburg-Landau equation (GLE) for the mesoscopic
superconducting thin film of the square shape in the magnetic field for the
wide range of the Ginzburg-Landau parameter $0.05<\kappa _{eff}<\infty $. We
focus on the region of the field where formation of the antivortex has been
reported previously. We found that the phase with the antivortex exists in
the broad range of parameters. When the coherence length decreases the
topological phase transition to the phase with the same total vorticity and
a reduced symmetry takes place. The giant vortex with the vorticity $m=3$ is
found to be unstable for any field, $\xi /a$ and $\kappa _{eff}\gtrsim 0.1$.
Reduction of $\kappa _{eff}$ does not make the phase with antivortex more
stable contrary to the case of the cylindric sample of the type I
superconductor.
\end{abstract}

\pacs{PACS: 74.60.Ec 74.25.Ha 74.80.-g}

\narrowtext
%\twocolumn

\section{ Introduction}

Recently it was shown that the influence of boundaries can lead to
stabilization of the vortex-antivortex molecules in mesoscopic samples\cite
{mosch}. Analysis of the linearized Ginzburg-Landau equation (GLE) has shown
that such molecules appear at particular values of the external magnetic
field depending on the sample shape and size\cite{chib}. The solution of the
GLE in the limit of the extreme type-II superconductor shows that such
molecules have a very shallow minimum in the free energy\cite{bonca,baelus}
and are very sensitive to the change of the sample shape\cite{meln}.

In a square mesoscopic thin film with the total vorticity $m=3$ the
symmetric solution with four vortices and one antivortex is the solution of
the linearized GLE with the lowest free energy\cite{mosch}. According to
ref. \cite{bonca}, away from the $H_{c2}$ line the giant vortex with
vorticity $m=3$ is stable and has the lowest free energy. This implies that
a topological phase transition {\em without change} of the vorticity and 
{\em without a reduction} of the symmetry should take place with the change
of the external field or/and the coherence length away from the
critical-field line.

It was proposed that in the limit $\kappa \simeq 1/\sqrt{2}$, where the
vortex-antivortex interaction changes the sign, vortex-antivortex complexes
should be more stable\cite{misko} for the cylindric sample shape. For a thin
film with $\lambda _{eff}=\lambda ^{2}/d$ and $\kappa _{eff}=\lambda
_{eff}/\xi $, where $d$ is the thickness of the film, $\lambda $ is the
London penetration depth, and $\xi $ is the superconducting coherence
length, this mechanism should be less effective due to a smaller
contribution of the 'magnetic energy' to the total free energy.

We performed an extensive study of the region of the phase diagram where the
vortex-antivortex phase was previously reported for the sample of the square
shape.\cite{mosch} We focused to the region $4<a/\xi <8$ and $\kappa
_{eff}>0.05$. We found that the antivortex phase is stable in a broad range
of parameters. The region of stability of the phase does not depend strongly
on the value of the parameter $\kappa _{eff}$. The energy gain due to the
antivortex formation is much smaller then the energy difference between two
phases with different vorticities. The giant vortex with $m=3$ is unstable
for any field, $\xi /a$ and $\kappa _{eff}\gtrsim 0.1$. Phase transition to
the phase with three separated vortices takes place when $\xi /a$ is driven
away from the critical field line. The reduction of $\kappa _{eff}$ does not
stabilize the antivortex phase for the thin film sample in the contrast to
the case of the cylindric sample ref.\cite{misko}.

\section{Formalism and Solution}

GLE for the normalized complex order parameter $\psi =\Psi /\Psi _{0}$, $%
\Psi _{0}=\sqrt{\beta /\left| \alpha \right| }$ has the following form: 
\begin{equation}
\xi ^{2}(i\nabla +{\frac{2\pi {\bf A}}{{\Phi _{0}}}})^{2}\psi -\psi +\psi
|\psi |^{2}=0
\end{equation}
here $\xi ={\frac{\hbar ^{2}}{{4m|\alpha |}}}$, $\alpha $ and $\beta $ are
the temperature dependent parameters of the Ginzburg-Landau expansion for
the free energy, $\Phi _{0}$ is the flux quantum, ${\bf A}$ is the vector
potential and ${\bf H}=\nabla \times {\bf A}$ the magnetic field. The second
GLE equation for the vector potential reads: 
\begin{equation}
\nabla \times \nabla \times {\bf A}=-i{\frac{\Phi _{0}}{{4\pi \lambda ^{2}}}}%
(\psi ^{*}\nabla \psi -\psi \nabla \psi ^{*})-{\frac{|\psi |^{2}{\bf A}}{{\
\lambda ^{2}}}}.
\end{equation}
In addition to Eq.(1) we assume the boundary condition for the
superconductor-insulator junction on the sample edges: 
\begin{equation}
(i\nabla +{\frac{2\pi {\bf A}}{{\Phi _{0}}}})\cdot {\bf n}\psi =0,
\end{equation}
where ${\bf n}$ is the vector normal to the surface of the sample.

As it was described in ref.\cite{bonca} we introduce $N\times N$ discrete
points on the square and rewrite Eq.(1) in the form of the nonlinear
discrete Schr\"{o}dinger equation: %\begin{equation}
%\sum_{{\bf l}}t_{{\bf i+l,i}}\psi _{{\bf i+l}}-4t_{{\bf i,i}}\psi _{{\bf i}
%}-\psi _{{\bf i}}+\psi _{{\bf i}}|\psi _{{\bf i}}|^{2}=0
%\end{equation}
\begin{equation}
\sum_{{\bf l}}t_{{\bf i+l,i}}\psi _{{\bf i+l}}-\epsilon ({\bf i})t_{{\bf i,i}%
}\psi _{{\bf i}}-\psi _{{\bf i}}+\psi _{{\bf i}}|\psi _{{\bf i}}|^{2}=0,
\end{equation}
where the summation index ${\bf l}=(\pm 1,0)$, $(0,\pm 1)$ points toward the
nearest neighbours and $t_{{\bf i_{1},i}}=(\xi N/a)^{2}exp(i\phi _{{\bf %
i_{1},i}})$ and $\phi _{{\bf i_{1},i}}=-{\frac{2\pi }{{\Phi _{0}}}}\int_{%
{\bf r}_{{\bf i}}}^{{\bf r}_{{\bf i_{1}}}}{\bf A}({\bf r})d{\bf r}$. The
boundary conditions are included in the discrete nonlinear Schr\"{ }odinger
equation as in ref.\cite{bonca} where $\psi _{{\bf i}}=0$ if ${\bf i}$ is
outside of the sample and $\epsilon ({\bf i})=4-\delta _{i_{x},1}-\delta
_{i_{x},N}-\delta _{i_{y},1}-\delta _{i_{y},N}$ where ${\bf i}%
=(i_{x}=1,\dots ,N,i_{y}=1,\dots ,N)$.

After discretization of Eq.(2) we can obtain the exact expression for the
vector potential: 
\begin{equation}
A_{{\bf i}}^{v}=\sum_{{\bf n}}K({\bf i-n})J_{{\bf n}}^{v}
\end{equation}
where 
\begin{eqnarray}
J_{{\bf i}}^{v} &=&\frac{\Phi _{0}a}{4\pi \lambda _{eff}N}\Im (exp(-i\phi _{%
{\bf i+l_{v},i}})\psi _{{\bf i}}^{*}\psi _{{\bf i+l_{v}}}-  \nonumber \\
&&exp(-i\phi _{{\bf i-l_{v},i}})\psi _{{\bf i}}^{*}\psi _{{\bf i-l_{v}}})
\end{eqnarray}
where $v\in \{x,y\}$ and ${\bf l_{x}}=(1,0)$, ${\bf l_{y}}=(0,1)$ 
\begin{equation}
K({\bf n})=\frac{N}{2\pi ^{2}a}\int_{0}^{\pi }dxdy\frac{%
cos(n_{x}x)cos(n_{y}y)}{\sqrt{4-2cos(x)-2cos(y)}}
\end{equation}

The numerical self consistent solution of the problem is obtained by
iterating the solution of the nonlinear equation for the order parameter
Eq.(4) and calculations of the current and the vector potential Eqs.(5,6).
We used two ways of solving Eq.(4). The first is similar to that reported in
ref.\cite{bonca} and corresponds to the iterative solution of the linearized
Eq.(4). The second relies on the fact that Eq.(4) represents the Euler
equation for the free-energy functional whith included boundary conditions.
Eq.(4) was therfore solved by the direct minimization of the corresponding
functional using the conjugate-gradient method. Both techniques gave
identical results.

\section{Results}

The main goal of the paper is to investigate the phase diagram in the region 
$4.5<\Phi /\Phi _{0}<6.5$ and $(a/\xi )^{2}<60$ where the solution with one
antivortex and four vortices (Fig.1) has been reported. We found that the
region of the phase diagram where the symmetry induced antivortex solution
has the lowest energy is broader than expected from the solution of the
linearized GLE. As it is shown in Fig. 2 for $\kappa _{eff}=\infty $ the
antivortex phase is stable up to $(a/\xi )^{2}\sim 55$, depending on $\Phi
/\Phi _{0}$. For a finite $\kappa _{eff}$ this region shifts to the higher
field as $(a/\xi )^{2}$ increases (see Fig.2).

The interesting behaviour is observed when the external field is fixed and $%
(a/\xi )^{2}$ increases. Close to the $H_{c2}$ the lowest minimum of the
free energy corresponds to the solution with the vorticity $m=4-1$ with the
antivortex in the center of the square. Present calculations do not confirm
the existence of the giant-vortex solution with $m=3$ in this region of the
phase diagram as reported previously \cite{bonca}. The difference is due to
increase of the number of discrete points $N$ enabling detection of the
antivortex. With increase of $(a/\xi )^{2}$ away from the $H_{c2}$ line the
phase transition to the multivortex state with the same vorticity ($m=3$)
and a lower symmetry takes place (see Fig. 2). In general, the free energy
depends on the vorticity $m=n_{+}-n_{-}$ and the total number of vortices in
the system $n=n_{+}+n_{-}$. The transition at $(a/\xi )^{2}\sim 55$ and $%
\Phi /\Phi _{0}\sim 5.5$ takes place at the constant vorticity $m=3$ with
the change of $n$ from 5 to 3. The transition is therfore not only
characterized by an order parameter, but also by the change of the number of
vortices at the constant total vorticity $m$, suggesting that the transition
is close to the first order. This statement is confirmed by the observation
that above the transition point, $(a/\xi )>(a/\xi )_{crit}$, both solutions
with $m=3$ and $m=4-1$ coexists (see Fig. 4). Since near the transition the
free-energy difference between the phases with the same vorticity $m$ and
different $n$ is small it is difficult to determine the phase boundary
between phases with $m=4-1$ and $m=3$ accurately. The transition could be
easier observed by calculating the two component order parameter $\eta
_{x}=\int x|\psi (x,y)|^{2}dxdy$, $\eta _{y}=\int y|\psi (x,y)|^{2}dxdy$
shown in Fig.3. The transition point is given by the point where the order
parameter goes to 0. At the same point the calculated sample magnetization
changes slope as it is clearly seen from Fig.4.

Close to the $H_{c2}$ line the repulsion of vortices from the boundaries and
attraction of the 4 vortices to the antivortex stabilizes the phase with $%
m=4-1$ and small vortex-vortex distances. At smaller value of $\xi $ the
repulsion from the boundaries decreases and one vortex annihilates with the
antivortex. As a result, the repulsion between the remaining vortices
increases leading to an increase of the order parameter with a further
decrease of $\xi$.

Increasing the field up to $\Phi /\Phi _{0}=10.6$ leads to the stabilization
of the phase with total vorticity $m=7$. Near the $H_{c2}$ line similar to
the phase with $m=4-1$ the solution with $m=8-1$ is realized. When $%
(a/\xi)^{2}$ increases in a complete analogy to the case with $m=4-1$ the
second order phase transition to the phase with $m=7$ and a similar order
parameter takes place. Here also, both solutions with $m=8-1$ and $m=7$
coexists above $(a/\xi)_{crit}$ indicating that the transition is close to
the first order. We believe that the situation is quite general for the case
of arbitrary $m=4l-1$ for $l=1,2,3,...$.

At the end we would like to discuss the dependence of the stability of the
antivortex phase at small $\kappa $. According to the arguments of Ref.\cite
{misko}, at small $\kappa $ the vortex-vortex interaction changes the sign
making the antivortex phase more stable. As a result, the average distance
between vortices in the middle of the square increases as well. In order to
verify this conjecture for the thin film sample we plot in Fig.5 the
vortex-antivortex distance $r_{0}$ as a function of $1/\kappa _{eff}$. The
distance decreases with the decreasing $\kappa _{eff}$. For $\kappa
_{eff}<0.1$ the distance is smaller than the grid spacing $a/N$ so we can
not resolve separate vortices. We find that $r_{0}\propto \exp {(-\Lambda
/\lambda _{eff})}$ {with }${\Lambda \sim a}$. The situation is just opposite
to that reported in ref.\cite{misko}. We believe that in the case of the
thin film of the square shape the reduction of $\kappa $ does not stabilize
the phase with the antivortex.

It is interesting to note differences between samples of different shapes.
For the cylindric shape the giant vortex phase with any vorticity is always
stable close to the $H_{c2}$ line\cite{Schw,Peet}. According to the Ref.\cite
{misko} for the mesoscopic triangle the giant vortex state with $m=2$ is
metastable and the solution with the antivortex ($m=3-1)$ is stable. For the
case of the square shape the giant vortex solution with $m=3$ is {\em never}
stable for $\kappa _{eff}\gtrsim 0.1$. For $\kappa _{eff}<0.1$ the limited
grid spatial resolution prevented us to distinguish the solution with the
antivotrex from the possibly (meta)stable giant-vortex solution.

Finally, let us discuss the possibility to detect the state with the
antivortex experimentally. Calculation of the magnetic field in the sample
shows that the magnetic field has a local minimum in the center of the
sample also for the giant vortex solution with $m=2$. The local minimum
observed for the antivortex state with $m=4-1$ is therfore not due to the
antivortex formation (Fig.6) but due to a particular distribution of the
current in the sample. Therefore, imaging of the magnetic field distribution
cannot provide an evidence for the antivortex. The magnetic field for the
multivortex solution with $m=3$ has 3 well separated maxima that break the
four fold rotational symmetry of the sample allowing a direct imaging of
vortices. Since the antivortex state cannot be detected directly the
observation of a hysteresis in the vicinity of the transition line from the $%
m=4-1$ antivortex state to the $m=3$ multivortex state could suggest that
the symmetric phase is indeed the phase with the antivortex.

{\Large Figure captions}

Fig. 1 (a) The magnitude of the superconductivity order parameter for $%
(a/\xi )^{2}=35$, $\kappa _{eff}=\infty $ and $\Phi =5.9\Phi _{0}$. (b) The
central region where the vortices are located in an expanded scale. The
position of the antivortex is indicated by the symbol $\otimes $.

Fig. 2 The calculated phase diagram. Different phases are marked with icons
schematically indicating the vortex pattern where the full dot represents a
vortex, the open dot reperesents an antivortex and the larger full dot
represents a double vortex. The full symbols and continous lines represent
the phase boundaries for $\kappa_{eff}=\infty$ while the open symbols and
dotted lines reperesent the phase boundaries for $\kappa_{eff}=1$. In the
later case only the phase boundaries of the region with the total vorticity
3 are shown.

Fig. 3 The magnitude of the order parameter around the transition where one
vortex anihilates with the antivortex as a function of $(a/\xi )^{2}$ at the
constant magnetic field.

Fig. 4 The magnetic moment of the sample as a function of $(a/\xi)^{2}$ at
the constant magnetic field.

Fig. 5 The vortex-antivortex distance (a) and the magnetic moment of the
sample (b) as functions of the parameter $1/\kappa _{eff}$. In (a) error
bars represent the grid spacing and the solid line the exponential fit
discussed in the text.

Fig.6 The magnetic field in the film in the case of (a) giant vortex with $%
m=2$, (b) the antivortex solution with $m=4-1$ and (c) three separate
vortices with $m=3$. Here $H_{0}$ is the external magnetic field.

\end{document}